# Data Accuracy Estimation for Cluster with Spatially Correlated Data in Wireless Sensor Networks

Jyotirmoy Karjee, H.S Jamadagni
Cente for Electronics Design and Technology
Indian Institute of Science
kjyotirmoy@cedt.iisc.ernet.in, hsjam@cedt.iisc.ernet.in

**Abstract-***Objective*-The main purpose of this paper is to construct a data accuracy model for the maximal set of sensor nodes that sense a point event and forms a cluster with fully connected network between them. We determine the minimal set of sensor nodes that are sufficient to give approximately the same data accuracy achieve by the maximal set of sensor nodes.
*Design approach/Procedure—L* set of sensor nodes are randomly deployed over a region *Z*. Since a point event *S* has occurred in the region *Z*, *M* maximal set of sensor nodes wake up and start sensing the point event. The set of *M* sensor nodes forms a cluster with fully connected network and remaining set of sensor nodes continue to be in sleep mode. One sensor node is elected randomly as a cluster head *(CH)* node which can estimate the data accuracy for the cluster before data aggregation and finally send the data to the sink node. *Findings* - Since we simulate the data accuracy for the cluster (*M* set of sensor nodes) at *CH* node, there exist *P* minimal set of sensor nodes which give approximately the same data accuracy level achieve by *M* set of sensor nodes .Moreover we find that as the distance from the point event to the number of sensor nodes increases, the data accuracy also get decreases. *Design Limitation* –This model is only applicable to estimate data accuracy for the point event where the sensed data are assumed to be spatially correlated with approximately same variations.
*Practical implementation*–Detect point event e.g. fire in forest. *Inventive/Novel idea* – This is the first time that a data accuracy model is performed for the cluster before data aggregation at the *CH* node which can reduce data redundancy and communication overhead.

***Keywords-Wireless sensor networks , Data accuracy ,Spatial correlation***

## I. INTRODUCTION

Recent development in wireless technology has made a drastic change in communication networks. One of the major tasks of wireless sensor networks is to sense or collect the physical phenomenon of data for the event such as fire, seismic event, temperature, humidity etc from the physical environment [1]. This physical phenomenon of data is sensed by a device called node where these raw data are proceed, communicated wirelessly and finally collected by sink node. Most of the data collected by sensor nodes are spatially correlated [2]. Since the data are spatially correlated in the sensor field, it is easier to estimate data accuracy at the sink node. Most of the work done till today is that the sink node or base station is responsible to estimate the data accuracy for the physically sensed data by the sensor nodes [3, 4, 5]. Hence this type of model is only suited for one hop communication i.e. raw data are sensed by the sensor nodes and directly transmitted to the sink node. But in this paper, we consider two hop communications where physical phenomenon of sensed data is transmitted via intermediate node called cluster head (*CH*) node. If we deploy sensor nodes randomly in a region and a point event has occurred, then a maximal set of sensor nodes starts sensing the point event and forms a cluster [12] with fully connected network. Remaining sensor nodes goes to be in sleep mode. One node is elected randomly by the cluster called *CH* node [11] which is responsible to perform data accuracy estimation for the cluster and finally these estimated data are retrieved by the sink node via *CH* node.

The main goal of this paper is to estimate data accuracy for the cluster (maximal set of sensor nodes) before data aggregation [13] at *CH* node which can reduce the data redundancy and communication overhead. However to the best understanding of the authors, there is no work done so far on verifying the data accuracy for cluster before data aggregation [15,16] at *CH* node. Since most of the done till today is that data from cluster of sensor nodes directly send to *CH* node for aggregation without verifying its accuracy. Hence it is important that the most accurate (precise or important) data send by the cluster of sensor nodes can aggregate at the *CH* node rather than aggregating all the redundant data at *CH* node. The data send by the cluster of sensor nodes should first verify its accuracy level at the *CH* node then only the data get aggregates and finally send to the sink node. Since *CH* node performs the data accuracy for the cluster, it can reduce the power consumption and may increase the lifetime of the networks. Another major importance to verify the estimated data accuracy for the cluster before data aggregation at *CH* node, if some of the sensor nodes in the cluster are malicious [14]. If the sensor nodes are malicious, it can sense and read inaccurate data. The inaccurate data send by the malicious node gets aggregated with the other correct data results in inaccurate (incorrect) data aggregation at the *CH* node and finally send to the sink node. This may increase the power consumption, data redundancy and communication overhead. It shows very high or low variations of the estimated data accuracy value compare to the actual variations of estimated data accuracy value at the *CH* node. Hence to overcome this problem, it is important to verify the data accuracy at *CH* node before data aggregation and send the accurate data to the sink node. Since in our assumptions the sensed data are spatially correlated with approximately the same variations and the sensor nodes are appropriate to sense the correct data, we get estimated data accuracy with approximately same variations at the *CH* node.

Hence in our model, we focus that a set of maximal sensor nodes which forms a cluster are responsible to sense the physical phenomenon of data such as temperature, humidity etc. Once the data accuracy is processed by *CH* node, it transmits the estimated accurate data to the sink node. From the literature survey, it is clear that only the sensor nodes are responsible to sense the physical phenomenon of data and not the sink node. But in our considerations not only sensor nodes are responsible to sense the physical phenomenon but the *CH* node can also do the sensing phenomenon. We investigate how the set of sensor nodes can sense the physical phenomenon of data to estimate the data accuracy. Literature [3, 6] has demonstrated some approaches regarding jointly sensing nodes which gives an idea about how the raw data is sensed by the jointly sensing nodes and how the number of jointly sensing nodes effects the data accuracy. However they address this problem if only sensing nodes are responsible to retrieve physical phenomenon of data where they investigate to find a proper number and positions of jointly sensing nodes. But in our model, we consider both the sensor nodes and the *CH* node which forms the cluster (maximal set of sensor nodes) are sensing the physical phenomenon such as temperature, humidity etc. Since we perform data accuracy for the cluster, there exit a minimal set of sensor nodes which give approximately the same data accuracy level achieve by same maximal set of sensor nodes.

The rest of the paper is given as follows. In the *section II*, we construct the system model where we defined how the sensor nodes are deployed and how to perform the normalized data accuracy for the maximal set of sensor nodes. In *section III* , we demonstrate spatial correlation model for normalized data accuracy and how on demand event detection can be done using spatial correlation model . In the *section IV,* we perform the simulation in grid topology and random topology for the maximal set of sensor nodes which forms the cluster. Moreover we also show how the distance from the point event to the number of sensor nodes effects the data accuracy. In *section V*, we discuss for the minimal set of sensor nodes which give approximately the same data accuracy achieves by the maximal set of nodes. Finally we concluded our work in *section VI*.

## II. SYSTEM MODEL

In this section, nodes deployment strategy is done in a sensor region and a mathematical foundation of data accuracy model is constructed for the single cluster in wireless sensor networks.

### A. *Sensor Nodes Deployment*

Let *L* be the set of sensor nodes which are randomly deployed over a region *Z* such that $Z \subseteq R^2$ where $||L||$ are the total number of sensor nodes. A point event *S* has occurred in the region Z. A point event *S* is an event that originates at a point in the region *Z* and radiates outwards e.g. fire. Suppose *M* be the maximal set of sensor nodes that sense the physical phenomenon such as temperature measurement for the point event *S* and forms a fully connected network between them as shown in Figure-1. A fully connected network is defined as a network topology in which there exists a direct one hop link between all pair of sensor nodes. If $||M||=m$ be the number of sensor nodes which wake up when a point event is sensed by them and there exists *m(m-1)/2* direct hop links to form a fully connected network with *m* sensor nodes. We assume that *M* set of sensor nodes form a cluster and one of the node is randomly elected as cluster head *(CH)* node [11] from the cluster *M* which aggregates all the physical phenomenon of data and send it to the sink node. Since our initial motivation is to construct the data accuracy model for cluster *M* to sense the actual point event *S* before all the data is being aggregated at *CH* node and send to the sink node. Hence we construct a mathematical model for data accuracy at the *CH* node which collects all the spatially correlated data sensed by the cluster *M*.

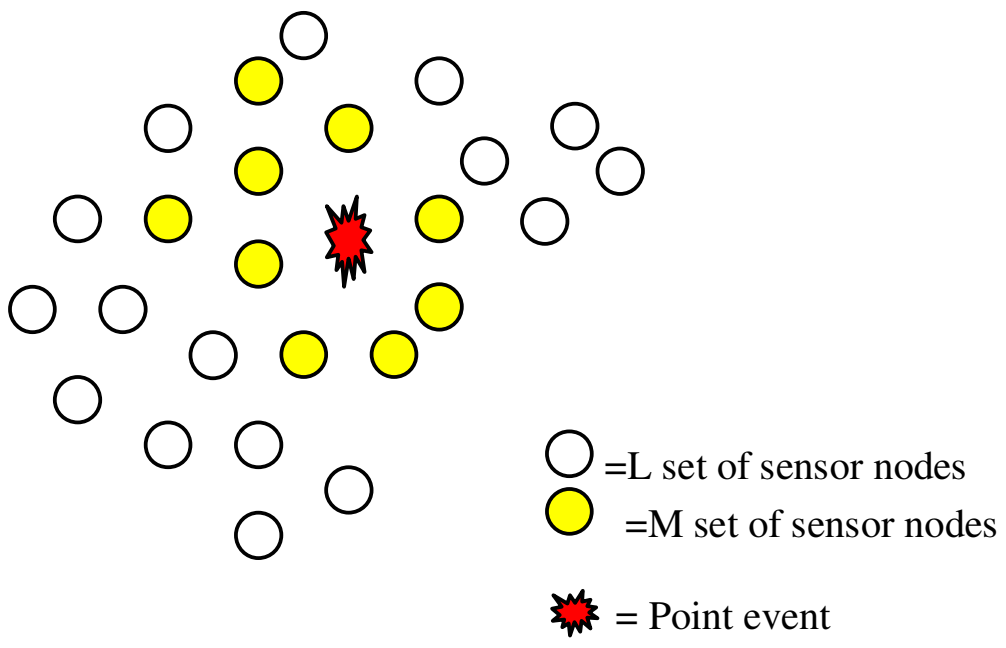

Figure 1: Sensor network topology

### B. *Mathematical Model for Cluster-based Data Accuracy*

Since *M* set of sensor nodes forms a cluster and can sense the physical phenomenon of data, we construct a methodology to find the data accuracy for the cluster before data aggregation at the *CH* node as shown in Figure-2.The data accuracy for the cluster is performed before data aggregation at the *CH* node to confirm that the data received at the *CH* node are accurate and not redundant which can reduce the communication overhead. Data accuracy is the degree of closeness of measurement for temperature of fire (point event) to its actual value. As we assume point event in our model, the initial assumption says that there are almost same variations in the physical phenomenon of data which is spatially correlated in the sensor region *Z*.

Notations used in data accuracy model for cluster are as follows:

- $S$= point event
- $\hat{S}$ = estimation of point event
- $S_i$ = physical phenomenon of $S$ sensed by node $i$ with no noise
- $\hat{S}_i$ = estimation of $S_i$
- $S_{CH}$ = physical phenomenon of S sensed by cluster head node with no noise
- $\hat{S}_{CH}$ =estimation of $S_{CH}$
- $X_i$=observed sample of $S_i$ by node $i$
- $Y_i$ = observed sample of $X_i$ under transmission noise
- $Z_i$=observed sample of $Y_i$ under power constraint
- $N_i$=noise under additive white Gaussian noise *(AWGN)*
- $N_{t_i}$ =transmission noise under *AWGN*
- $||M||=m$= total number of wake up nodes

- $d_{S,i}$= distance between $S$ and node $i$
- $d_{S,CH}$ = distance between $S$ and $CH$ node
- $d_{CH,i}$ = distance between $CH$ node and node $i$
- $d_{i,j}$ =distance between nodes $i$ and $j$

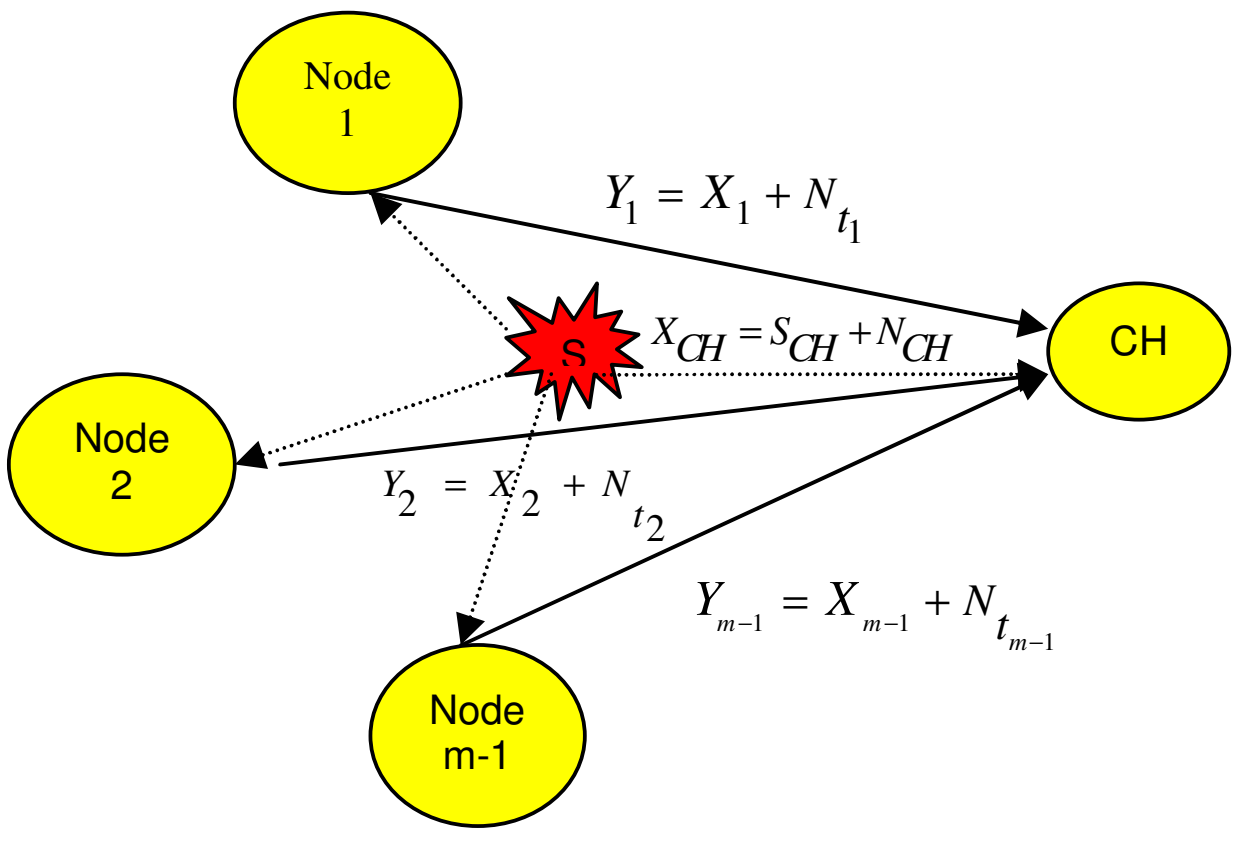

Figure 2: Cluster –based Data Accuracy Model

Each sensor node $i$ (where $i \in M$) in the cluster $M$ can observe the physically sensed data $S_i$ for point event $S$ with observation noise $N_i$. Hence the observation made by the sensor node $i$ is given by

$$X_i = S_i + N_i \qquad where\ i \in M \tag{1}$$

Since the sensor node $i$ (where $i \in M$ and $i \neq CH$) sense the observe sample $X_i$, it transmits $X_i$ to cluster head node sharing wireless additive white Gaussian noise (AWGN) channel [3,7]. Hence the observation received by the $CH$ node from other sensor nodes in the network with transmission noise $N_{t_i}$ over the AWGN channel represented as

$$Y_i = X_i + N_{t_i} = S_i + N_i + N_{t_i} \qquad where\ i \in M\ and\ i \neq CH \tag{2}$$

Since uncoded transmission with finite number of sensor nodes which is optimal for point-to-point transmission [4] and adopt the encoding power constraint value $P$ , the observed value received by the $CH$ are scaled as given bellow

$$Z_i = \sqrt{\frac{P}{(\sigma_{S_i}^2 + \sigma_{N_i}^2 + \sigma_{N_{t_i}}^2)}} Y_i = \alpha(S_i + N_i + N_{t_i}) \qquad where\ i \in M\ and\ i \neq CH \tag{3}$$

and $$\alpha = \sqrt{\frac{P}{(\sigma_{S_i}^2 + \sigma_{N_i}^2 + \sigma_{N_{t_i}}^2)}}$$

$CH$ node estimate the event source $S$ by calculating the estimation of each physical phenomenon $S_i$ for node $i$. We adopt minimum mean square estimation (MMSE) for optimal decoding phenomenon [8] for uncoded transmission .Finally $CH$ node calculate the MMSE for physical phenomenon $S_i$ extracted by sensor node $i$ with observation sample $Z_i$ given by

$$\hat{S}_i = \frac{E[S_i Z_i]}{E[Z_i^2]} Z_i \qquad where\ i \in M\ and\ i \neq CH \tag{4}$$

Since the sensor node $i$ sense physical phenomenon $S_i$ of $S$ , we assume independent identically distributed (i.i.d) Gaussian random variable with zero mean and variance $\sigma_S^2$ i.e E[S]=$0$ , var[S]= $\sigma_S^2$ for event source . Similarly for sensing phenomenon $S_i$, we adopt E[$S_i$]=$0$ , var[$S_i$]= $\sigma_{S_i}^2$ . We also represent the observation noise $N_i$ and transmission noise $N_{t_i}$ with an independent identically distributed Gaussian random variable with variances $\sigma_{N_i}^2, \sigma_{N_{t_i}}^2$ respectively where means are zero.

Hence E[$N_i$]=0 , E[ $N_{t_i}$ ]=0,var[$N_i$]= $\sigma_{N_i}^2$ ,var[ $N_{t_i}$ ]= $\sigma_{N_{t_i}}^2$ respectively.

Thus,
$$E[S_i Z_i] = \alpha \sigma_{S_i}^2$$
$$E[Z_i^2] = \alpha^2 (\sigma_{S_i}^2 + \sigma_{N_i}^2 + \sigma_{N_{t_i}}^2)$$

Thus the estimation of $\hat{S}_i$ is given by

$$\hat{S}_i = \frac{\sigma_{S_i}^2}{(\sigma_{S_i}^2 + \sigma_{N_i}^2 + \sigma_{N_{t_i}}^2)} (S_i + N_i + N_{t_i}) \tag{5}$$

where $i \in$ M and $i \neq$ CH

where $$\beta_i = \frac{\sigma_{S_i}^2}{(\sigma_{S_i}^2 + \sigma_{N_i}^2 + \sigma_{N_{t_i}}^2)} \qquad \text{for } 0<\beta_i<1 \tag{6}$$

It is clear that $CH$ node also performs the sensing phenomenon independent of all the sensor nodes with out any transmission noise. Since the $CH$ node also sensed the physical phenomenon $S_{CH}$ of event source $S$ and doesn't require the uncoded transmission for optimal decoding scheme with power constraint $P$, it simply calculate the MMSE for physical phenomenon $S_{CH}$ from the observation $X_{CH}$ *(where* $X_{CH}$=$S_{CH}$ + $N_{CH}$*)* follows

$$\hat{S}_{CH} = \frac{E[S_{CH} X_{CH}]}{E[X_{CH}^2]} X_{CH} \tag{7}$$

The observation noise $N_{CH}$ of $CH$ node can be represented as i.i.d Gaussian random variable with zero mean and variance $\sigma_{N_{CH}}^2$ we get

$$E[S_{CH} X_{CH}] = \sigma_{S_{CH}}^2$$

$$E[X_{CH}^2] = (\sigma_{S_{CH}}^2 + \sigma_{N_{CH}}^2)$$

Thus the estimation of $\hat{S}_{CH}$ is given by

$$\hat{S}_{CH} = \frac{\sigma^2_{S_{CH}}}{(\sigma^2_{S_{CH}} + \sigma^2_{N_{CH}})}(S_{CH} + N_{CH}) \qquad (8)$$

$$\text{where} \quad \beta_{CH} = \frac{\sigma^2_{S_{CH}}}{(\sigma^2_{S_{CH}} + \sigma^2_{N_{CH}})} \quad \text{for } 0< \beta_{CH} <1 \qquad (9)$$

From *equations no (6) and (9),* we get two constraint factors $\beta_i$ and $\beta_{CH}$ which controls data accuracy under Gaussian noise. Hence *M* set of sensor nodes forms a cluster and perform the sensing phenomenon such as temperature measurement when a point event (fire) has occurred in the region *Z*. We measure the data accuracy performance before aggregation of data at the *CH* node for *M* set of sensor nodes to sense the point event. To calculate the estimate of point event *S* at the *CH* node*,* we compute the average of the entire MMSE observation sample done by *m* sensor nodes and the expression for average estimate is given by

$$\hat{S}(M) = \frac{1}{m}\left[\sum_{i=1}^{m-1} \beta_i (S_i + N_i + N_{t_i}) + \beta_{CH}(S_{CH} + N_{CH})\right] \qquad (10)$$

The data accuracy $D(M)$ for the estimations is defined in terms of the expectation of the error between the actual value of point event and the mean square average estimates value of *M* set of sensor nodes. Hence we adopt mean square error between *S* and $\hat{S}(M)$ to verify the data accuracy estimation for the cluster which is given by

$$D(M) = E[(S - \hat{S}(M))^2]$$

$$D(M) = E[S^2] - 2E[S\hat{S}(M)] + E[\hat{S}(M)^2] \qquad (11)$$

The normalized [2] data accuracy $D_A(M)$ is given by

$$D_A(M) = 1 - \frac{D(M)}{E[S^2]}$$

$$D_A(M) = \frac{1}{E[S^2]}[2E[S\hat{S}(M)] - E[\hat{S}(M)^2]] \qquad (12)$$

The normalized data accuracy can be implemented in spatial correlation model explained in the next part.

## III. Spatially Correlated Data Accuracy Model

In this section, a spatial correlation model is constructed for normalized data accuracy for the cluster. Moreover we have a theoretical demonstration and implementation of this model for on-demand event detection.

### A. *Spatial Correlation Model*

Here we derive a mathematical model where all the sensed data are spatially correlated among them. These spatial correlations among data are achieved by *M* set of sensor nodes .We model spatially correlated physical phenomenon of sensed data as joint Gaussian random variables (JGRV's) [5] as follows:

Step 1: $E[S] = 0$ , $E[S_i] = 0$ , $E[S_{CH}] = 0$ ;
$E[N_i] = 0$ , $E[N_{t_i}] = 0$, $E[N_{CH}] = 0$

*Step 2:* $var[S] = \sigma^2_S$ , $var[S_i] = \sigma^2_{S_i}$ , $var[S_{CH}] = \sigma^2_{S_{CH}}$
$var[N_i] = \sigma^2_{N_i}$ , $var[N_{t_i}] = \sigma^2_{N_{t_i}}$ , $var[N_{CH}] = \sigma^2_{N_{CH}}$

*Step 3:* $\text{cov}[S, S_i] = \sigma^2_S corr[S, S_i]$
$\text{cov}[S, S_{CH}] = \sigma^2_S corr[S, S_{CH}]$
$\text{cov}[S_i, S_j] = \sigma^2_S corr[S_i, S_j]$
$\text{cov}[S_{CH}, S_i] = \sigma^2_S corr[S_{CH}, S_i]$

*Step 4:* $E[S, S_i] = \sigma^2_S corr[S, S_i] = \sigma^2_S \rho(s, i) = \sigma^2_S K_V(d_{s,i})$
$E[S, S_{CH}] = \sigma^2_S corr[S, S_{CH}] = \sigma^2_S \rho(s, CH) = \sigma^2_S K_V(d_{S,CH})$
$E[S_i, S_j] = \sigma^2_S corr[S_i, S_j] = \sigma^2_S \rho(i, j) = \sigma^2_S K_V(d_{i,j})$
$E[S_{CH}, S_i] = \sigma^2_S corr[S_{CH}, S_i] = \sigma^2_S \rho(CH, i) = \sigma^2_S K_V(d_{CH,i})$

Illustration of step *1-2* is already explained in the *section- II (B)*. Using step *3-4,* we explain the covariance model [9] for spatially correlated data. To clarify the covariance model say $\text{cov}[S_i, S_j] = E[S_i, S_j] = \sigma^2_S corr[S_i, S_j] = \sigma^2_S \rho(i, j) = \sigma^2_S K_V(d_{i,j})$ where $d_{ij} = \| S_i - S_j \|$ represents the Euclidean distance between node $n_i$ and $n_j$ and $K_V(.)$ is the correlation model for spatially correlated data. The covariance function is non-negative and decrease monotonically with the Euclidean distance $d_{ij} = \| S_i - S_j \|$ with limiting values of 1 at *d=0* and of 0 at $d=\infty$. We have chosen power exponential model [10] i.e. $K_V^{P.E}(d_{i,j}) = e^{-(d_{i,j}/\theta_1)^{\theta_2}}$ , $\theta_1 > 0; \theta_2 \in (0,2]$ $\theta_1$ is called a '*Range parameter*' which controls how fast the spatially correlated data decays with the distance. $\theta_2$ is called a '*Smoothness parameter*' which controls the geometrical properties of wireless sensor field.

Using (1),*(2)*and *(10)* in *(12)*, we interpret and verify the normalized data accuracy with spatial correlation model for the cluster (*M* set of sensor nodes ) as follows:

$$D_A(M) = \frac{1}{m}\left[\beta_i (2\sum_{i=1}^{m-1} e^{(-d_{(S,i)}/\theta_1)^{\theta_2}} - 1) + 2\beta_{CH} e^{(-d_{(S,CH)}/\theta_1)^{\theta_2}}\right]$$
$$- \frac{1}{m^2}\left[\beta_i (\beta_i \sum_{i=1}^{m-1}\sum_{j \neq i}^{m-1} e^{(-d_{(i,j)}/\theta_1)^{\theta_2}} - 1) + \beta_{CH}(2\beta_i \sum_{i=1}^{m-1} e^{(-d_{(CH,i)}/\theta_1)^{\theta_2}} + \beta_{CH})\right] \qquad (13)$$

The *equation-13* shows that the normalized data accuracy $D_A(M)$ depends upon *m* sensor nodes and factors $\beta_i$ and $\beta_{CH}$ respectively.

### B. *On Demand Event Detection for Spatial Correlation Model*

The motivation of this paper is to estimate the data accuracy level for the cluster of sensor nodes before aggregating the most appropriate (or accurate) data at the *CH* node. Hence it is necessary for the data send by the cluster of sensor nodes should first verify its accuracy level at the *CH* node then only the most accurate data get aggregated and finally send to the sink node. In a practical scenario, suppose *L* set of sensor nodes are deployed in a forest and a point event (e.g. fire) has occurred. Once a point event is detected, *M* set of sensor nodes starts sensing the physical phenomenon of data (e.g. temperature measurement) which forms a cluster with fully connected network between them and report to the sink node as fast as possible via *CH* node.

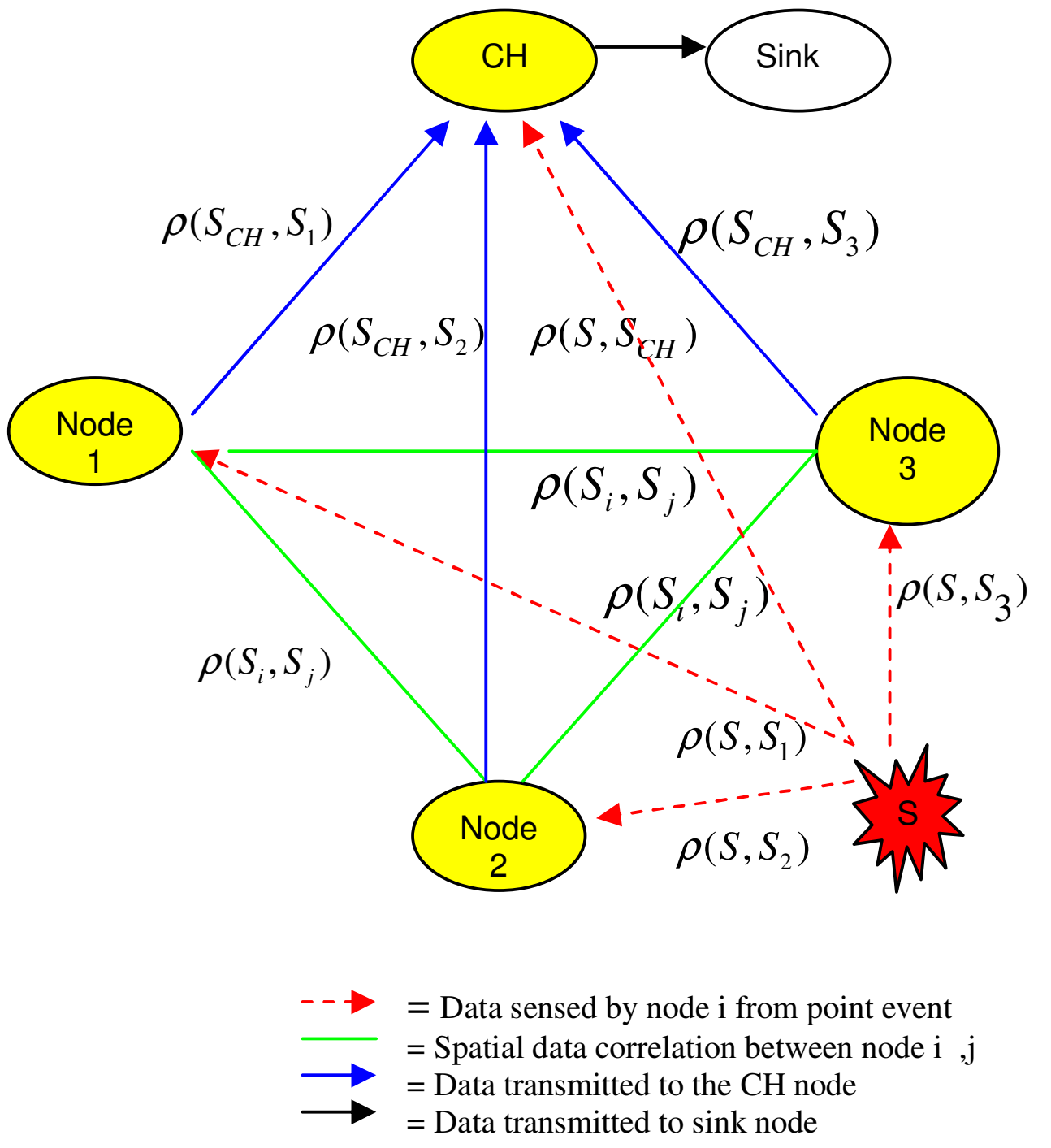

Figure 3: Spatial Correlation Model

Hence it is necessary that the most accurate (or important) data send by *M* set of sensor nodes can aggregated at the *CH* node rather than aggregating all the redundant data which can reduce the power consumption and communication overhead of the network .Since we get a normalized data accuracy at the *CH* node for *M* set of sensor nodes, we construct a spatial correlation model for spatially correlated data for wireless sensor network as given in *equation-13*. The spatial correlation model can be defined as:

- Each sensor node *i* can sense a point event *S* where $i \in M$ and $i \neq CH$ node
- *CH* node itself can sense the point event S.
- A spatial data correlation between node *i, j* where $i,j \neq CH$ node.
- Each sensor node *i* transmits the sensed data to the *CH* node where $i \in M$ and $i \neq CH$.

To visualize the spatial correlation model, we take an example where *m=4* sensor nodes and out of *m* sensor nodes one node is chosen as a *CH* node as shown in Figure-3.Once we estimate the data accuracy at *CH* node for the cluster , the most accurate data get aggregated and finally send to the sink node.

## IV. Simulation Results

We perform simulations to verify $D_A(M)$ for the cluster before data aggregation at *CH* node which depends on *m* sensor nodes and the factors $\beta_i$ and $\beta_{CH}$ respectively. We assume *L* set of sensor nodes are deployed in the region *Z*. *M* set of sensor node wake up randomly from *L* set of sensor nodes when a point event is detected and forms a cluster with fully connected network. In the first simulation set up, suppose *m=4* sensor nodes can sense a point event and forms a cluster with fully connected network. We put *m* sensor nodes in deployed circle and a point event S occurred at the centre of the deployed circle. i.e $d_{S,i}$ *(where i=1,2,3)*and $d_{S,CH}$ are equidistance as shown in the Figure-4 . In this case we have fixed the number of *m* sensor nodes and vary the distance from the point event *S* to *m* sensor nodes. As we increase the radius of the deployed circle for $d_{S,i}$ and $d_{S,CH}$ with same proportion , $D_A(M)$ decreases i.e. the distance from the point event *S* to the *m* sensor nodes increases as shown in Figure-5. We choose $\theta_1 = \{50,100\}$ and $\theta_2 = 1$ for our statistical data to perform the normalized data accuracy $D_A(M)$.

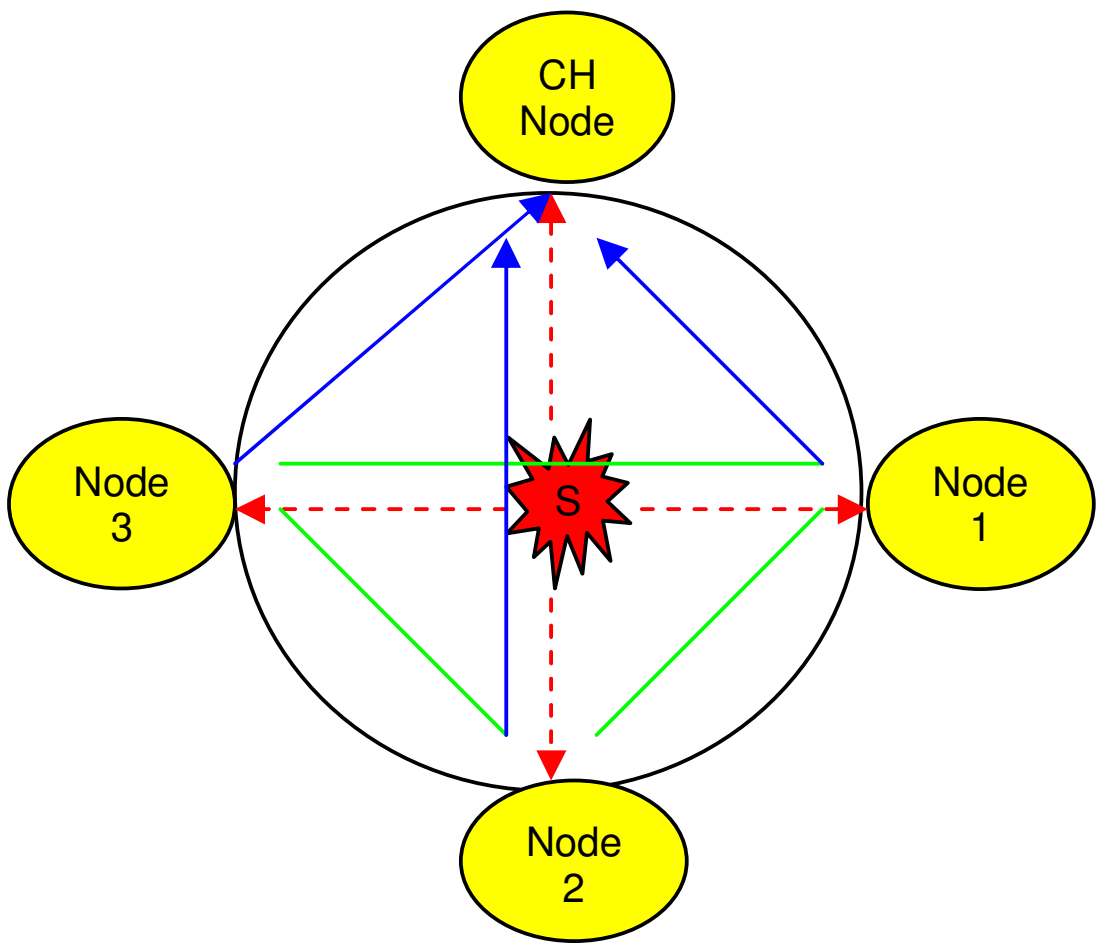

Figure 4: Deployed sensor nodes in circular topology

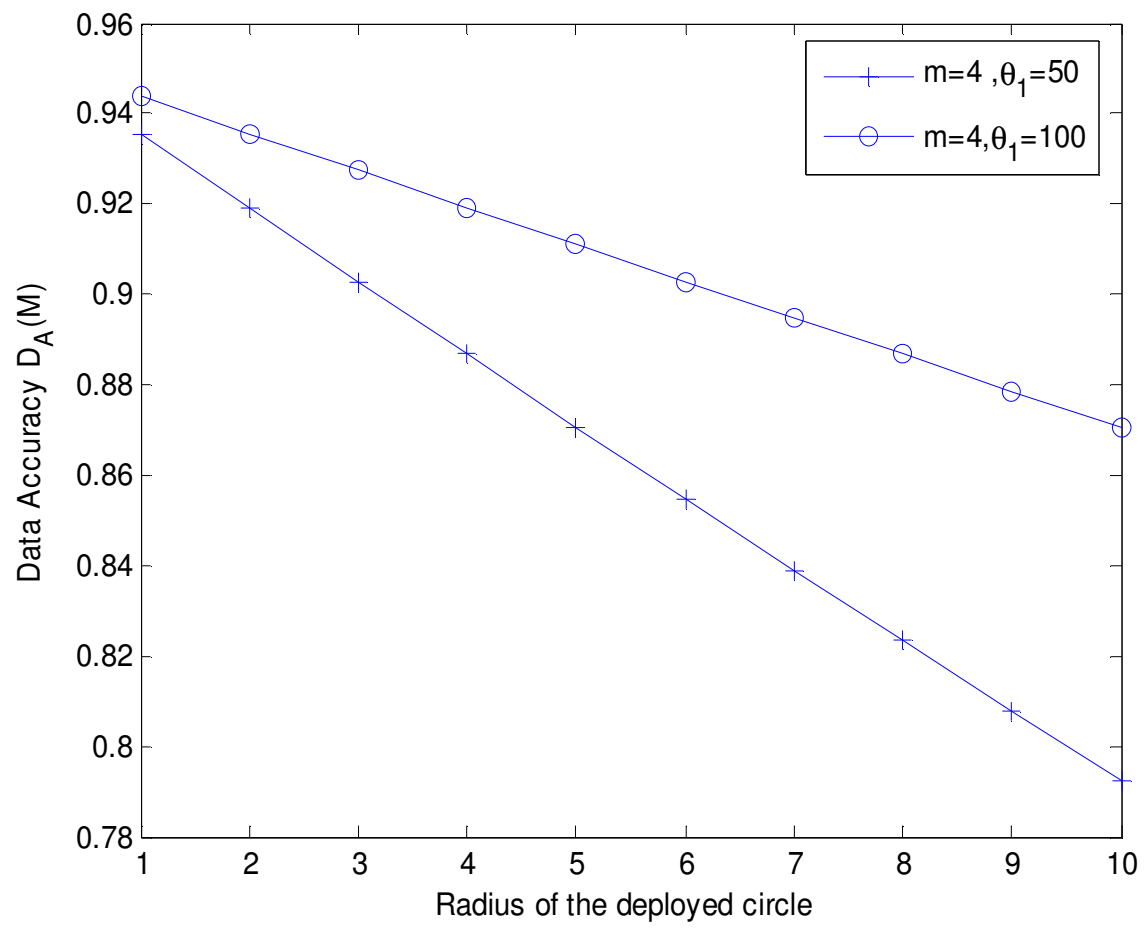

Figure 5: Data accuracy versus radius of the deployed circle

In the second simulation setup, the distance from the point event *S* to *m* sensor nodes is fixed in the deployed circle of radius =5 metre. We increase the number of sensor nodes with a fixed distance from the point event *S* i.e we increase *m* sensor nodes with fixed deployed circle of radius 5 metre. Initially we put *m=2* (one *CH* node and one sensor node) where the data accuracy is very poor with value in between 0.6 to 0.75 for $\theta_1 = \{50,100,200,400\}$.This is because there is only one sensor node which clarify that the third condition of spatial correlation model given in *section III(B)* doesn't satisfies the $D_A(M)$ at the CH node . But if *M=3* (one cluster head and two sensor nodes), there is a drastic improvement over $D_A(M)$ as all the conditions for spatial correlation model are satisfied. The Figure-6 also shows that five to eight nodes are sufficient to perform the $D_A(M)$ , if the distance from point event to *m* sensor nodes with deployed circle of radius is 5 metre.

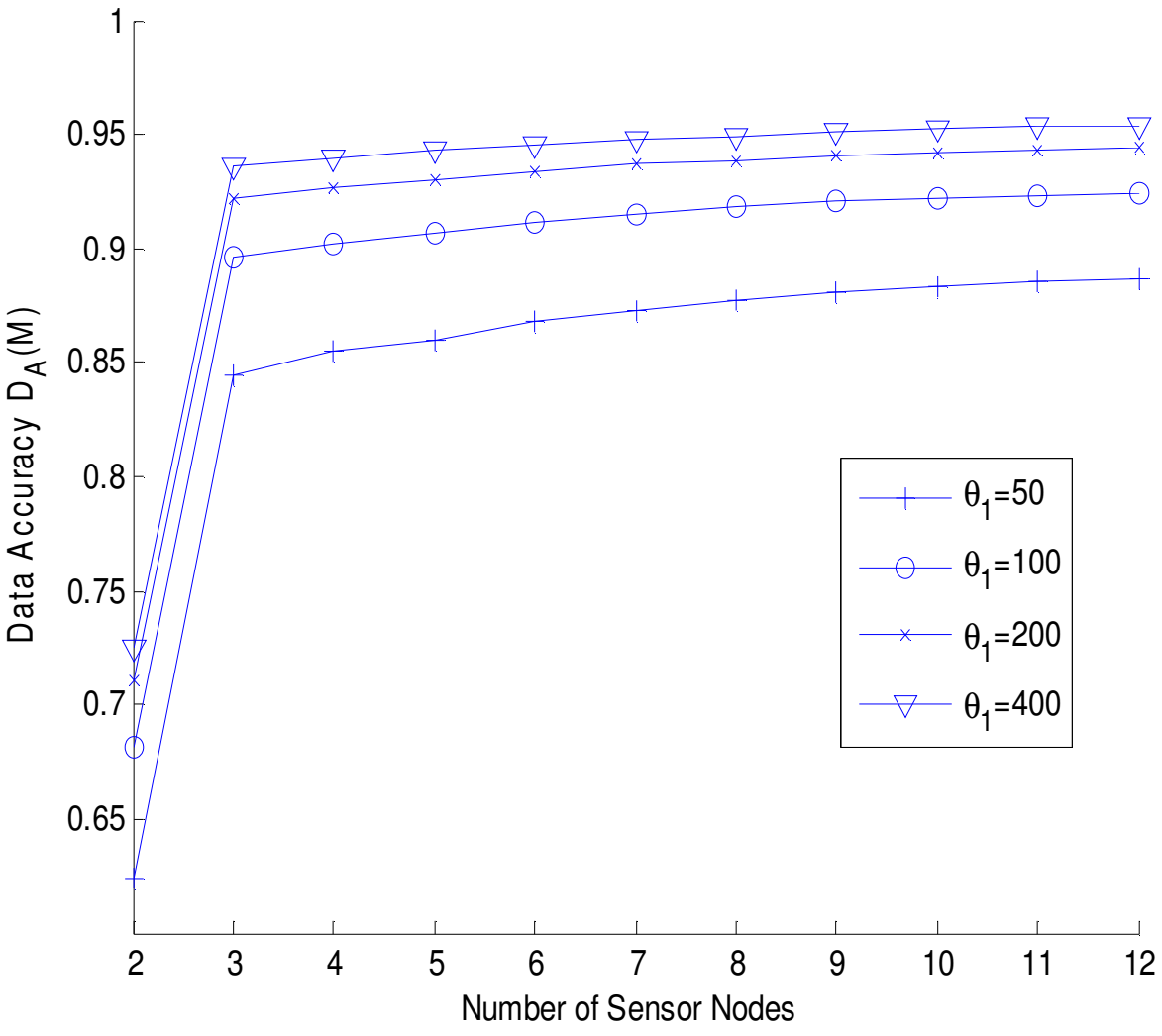

Figure 6: Data accuracy versus number of sensor nodes

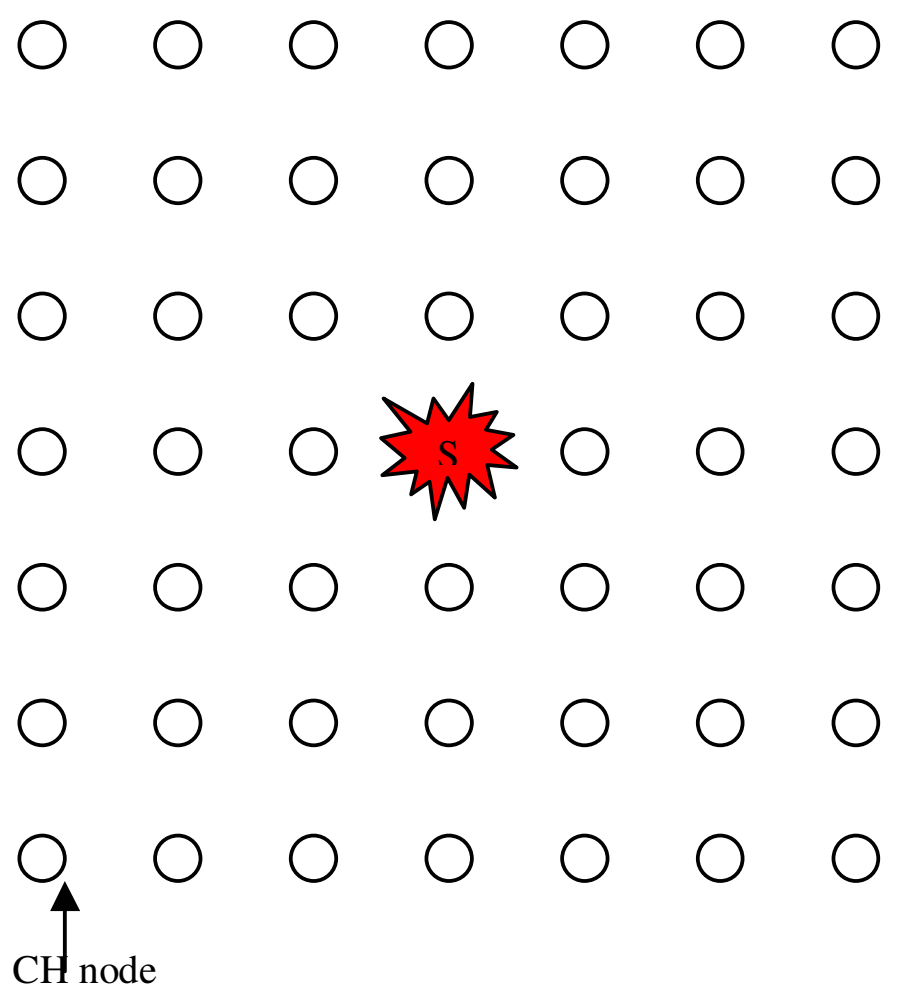

Figure 7: Sensor nodes deployed in grid topology

In the third simulation setup, we have simulated a wireless sensor field (900 metre$^2$) of 5m X 5m grid based sensor topology with a fixed event source *(S)* at the centre and a *CH* node on the corner edge with 47 sensor nodes distributed uniformly in the grid based sensor topology as shown in Figure-7.Our assumptions is that *m* sensor nodes are in the sensing range of the point event *(S)* and form a cluster with fully connected network. Initially we have chosen *m=4*(one cluster head node and three sensor nodes located at the four extreme corner of sensor field).We analyze that $D_A(M=4)$ is 0.6333 when $\theta_1$ =50 as shown in Figure -8. If we increase $\theta_1$ = 400, then $D_A(m=4)$ =0.911 which clarify that $\theta_1$ controls how fast the spatially correlated data decays with distance between sensor nodes and the event source. Hence it shows that it is always appreciable to take the value of $\theta_1$ large for large sensor field to get $D_A(M)$ in an efficient way. Now we increase *m* sensor nodes with increment of four sensor nodes every time concentrating towards event source till *m* sensor nodes are able to sense the point event *S* in the region. As we increase the sensor nodes, the data accuracy $D_A(M)$ also get increases. Hence for 900 metre$^2$ sensor field, 15 to 20 sensor nodes are sufficient to give $D_A(M)$ of 0.944 for $\theta_1$ =400 and $D_A(M)$ remains approximately constant still we increase the number of sensor nodes. We plot in the Figure-8 for the $D_A(M)$ versus node density. Node density is defined as the number of sensor nodes per unit area .Hence it is unnecessary to choose so many sensor nodes to achieve data accuracy for the cluster in sensor field to sense a point event.

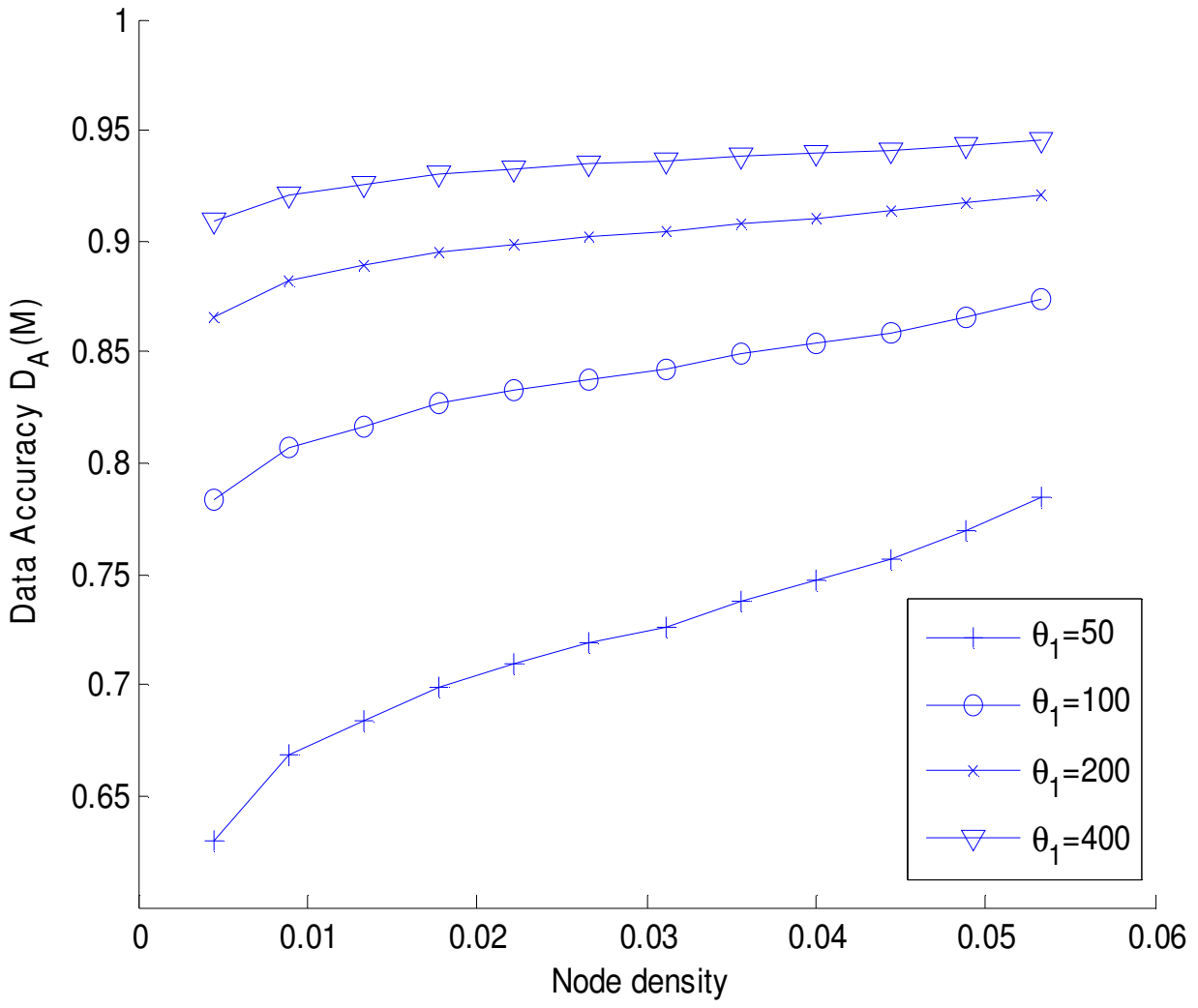

Figure 8: Data accuracy vs. node density

In the fourth simulation setup, *m* sensor nodes are randomly deployed in a region (30 X 30 = 900 metre$^2$) that sense a point event and forms a fully connected network between them. We fix the point event at x,y (15,15) coordinate and *CH* node at x,y (0,0) coordinate with 99 sensor nodes randomly deployed in the region. For each run we perform $D_A(M)$ with respect to randomly deployed *m* sensor nodes. Finally we perform 100 runs and find the average $D_A(M)$ for *m* sensor nodes. Figure-9 shows that if $\theta_1$ =400, $D_A(M)$ is 0.944 for 10 to 15 sensor nodes. If we increase the number of sensor nodes the $D_A(M)$ remains approximately same. Hence it is unnecessary to deploy sensor nodes beyond 15 sensor nodes because 10 to 15 sensor nodes are sufficient to give approximately the same $D_A(M)$ for $\theta_1$ =400. More over as $\theta_1$ increases, average $D_A(M)$ also increases for *m* sensor nodes. But after a certain approximate value of $\theta_1$ the $D_A(M)$ remains approximately same. If we continuously increase the value of $\theta_1$ the average $D_A(M)$ remains approximately constant since it achieve the saturation level. Finally the graph shows distortion in the output because additive white Gaussian noise components are embedded in the signal.

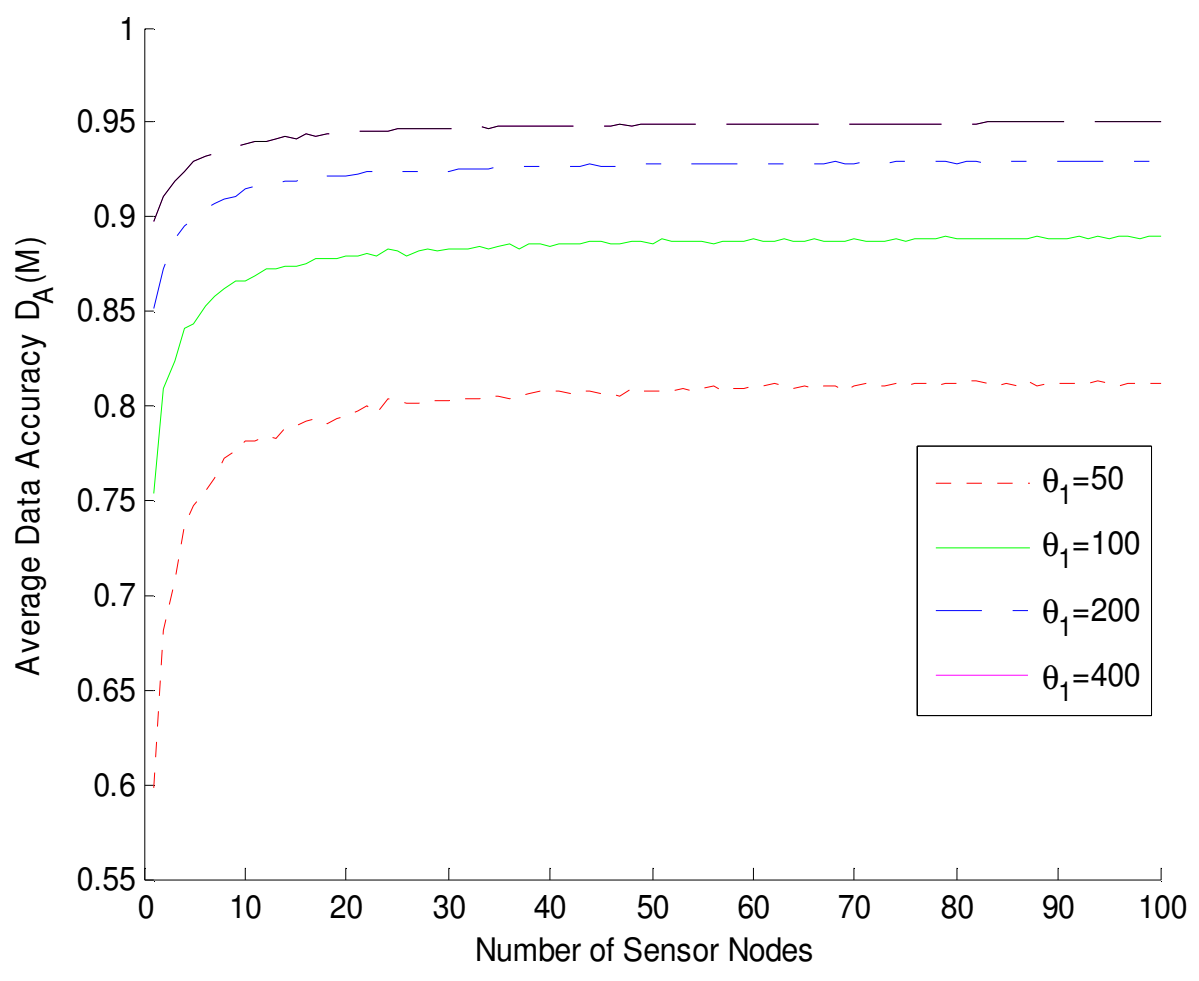

Figure 9: Average data accuracy versus number of sensor nodes

## V. Data Accuracy for Minimal Set of Sensor Nodes

Since initially we deployed *L* set of sensor nodes in a region *Z*. Out of *L* sensor nodes, *M* set of sensor nodes wakeup when a point event is detected and forms a cluster with fully connected network. We perform $D_A(M)$ for the cluster before data aggregation at *CH* node. Hence measuring the most accurate data send by the cluster of sensor nodes can aggregate at the *CH* node rather than aggregating all the redundant data at the *CH* node .It can reduce the data redundancy. Once the $D_A(M)$ is performed for the cluster, it is clear from the simulation results that their exist *P* set of sensor nodes which forms an optimal cluster are sufficient to give approximately the same $D_A(M)$. Hence the time complexity calculated at the *CH* node for aggregating the most accurate data send by the optimal cluster *P* will be less. Moreover the power consumption for aggregating all the data for *P* optimal cluster at *CH* node will also be less. Thus the optimal cluster *P* can reduce the data redundancy and communication overhead at the *CH* node.

In the third simulation, we deployed *m=48* sensor nodes in a grid topology and examine that 15 to 20 nodes are sufficient to perform $D_A(M)$ =0.944 for $\theta_1$ =400 in 900 metre$^2$ region. Similarly in the fourth simulation, we deployed *m=100* sensor nodes randomly in 900 metre$^2$ region and find that 10 to 15 sensor nodes are sufficient to perform $D_A(M)$ =0.944 for $\theta_1$ =400. Hence it is unnecessary to choose so many sensor nodes in 900 metre$^2$ region as $D_A(M)$ remains approximately constant as it achieve the saturation level still we increase *m* sensor nodes . Hence we define *P* minimal set of sensor nodes which forms an optimal cluster are sufficient to give approximately the same $D_A(M)$ by *M* maximal set of nodes as shown in Figure-10. The remaining set of nodes continues to be in sleep mode.

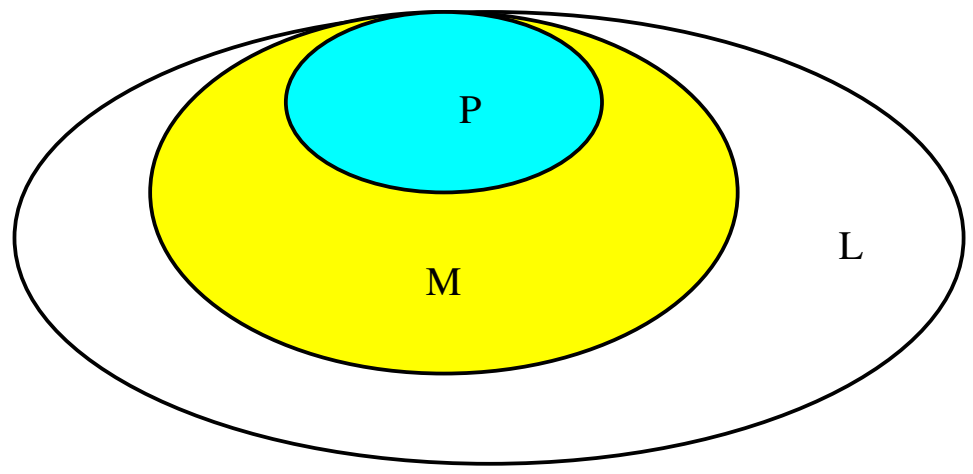

Figure 10: Venn diagram for deployed sensor nodes

## VI. Conclusions

In this paper, we focus a deployment strategy of *L* set of sensor nodes in a region Z where out of *L* sensor nodes, *M* set of sensor nodes forms a cluster which are responsible for sensing a point event and estimate the data accuracy for the cluster before data aggregation at *CH* node. The most accurate data send by the cluster of sensor nodes can aggregate at the *CH* node rather than aggregating all the redundant data at *CH* node. We also stated that two constraint factors $\beta_i$ and $\beta_{CH}$ controls data accuracy under additive white Gaussian noise. We conclude that data accuracy for the cluster depends on number of sensor nodes. We perform simulations in grid topology as well as in random topology to estimate the data accuracy for the *M* deployed set of sensor nodes. Our simulation results shows that *P* (optimal cluster) minimal set of sensor nodes are adequate to sense the physical phenomenon of data for a point event to perform approximately the same data accuracy level achieve by *M* set of sensor nodes and also clarifies that as the distance increases from the point event to the sensing nodes data accuracy decreases. Finally we conclude that data accuracy performed for the cluster before data aggregation at *CH* node can reduce the data redundancy and communication overhead.